\documentclass[12pt]{article}

\usepackage{scicite}

\usepackage{bm}% bold math
\usepackage{graphicx}% Include figure files
\usepackage{times}

\newenvironment{sciabstract}{%
\begin{quote} \bf}
{\end{quote}}

\renewcommand\b[1]{{\bf  #1}}
\renewcommand\vec[1]{\boldsymbol{#1}}

\newcommand\tr{\mathrm{tr}}
\newcommand\del{\nabla}

\newcommand\dd{\mathrm{d}}

\title{Geometric control of topological dynamics in a singing saw}

\author
{Suraj Shankar$^{1\dagger}$, Petur Bryde$^{2\dagger}$, L.~Mahadevan$^{1,2,3\ast}$\\
\\
\small{$^{1}$Department of Physics, Harvard University, Cambridge, MA 02138, USA}\\
\small{$^{2}$Paulson School of Engineering and Applied Sciences, Harvard University, Cambridge, MA 02138, USA}\\
\small{$^{3}$Department of Organismic and Evolutionary Biology, Harvard University, Cambridge, MA 02138, USA}\\
\\
\small{$^\dagger$These authors contributed equally to this work.}\\
\small{$^\ast$Correspondence to:~lmahadev@g.harvard.edu}
}

\date{}

\begin{document} 


\baselineskip24pt


\maketitle


\begin{sciabstract}
The common handsaw can be converted into a bowed musical instrument capable of producing exquisitely sustained notes when its blade is appropriately bent. Acoustic modes localized at an inflection point are known to underlie the saw's sonorous quality, yet the origin of localization has remained mysterious. Here we uncover a topological basis for the existence of localized modes, that  relies on and is protected by spatial curvature. By combining experimental demonstrations, theory and computation, we show how spatial variations in blade curvature control the localization of these trapped states, allowing the saw to function as a geometrically tunable high quality oscillator. Our work establishes an unexpected connection between the dynamics of thin shells and topological insulators, and offers a robust principle to design high quality resonators across scales, from macroscopic instruments to nanoscale devices, simply through geometry.
\end{sciabstract}

Musical instruments, even those made from everyday objects such as sticks, saws, pans and bowls \cite{fletcher}, must have the ability to create sustained notes for them to be effective. While this ability is often built into the design of the instruments, the musical saw, used to make music across the world for over a century and a half~\cite{Sawmusical}, is unusual in that it is just a carpenter's saw, but held in an unconventional manner to allow it to sing.  When a  saw (Fig.~1A) is either bowed or struck by a mallet,  it produces a sustained sound that mimics a ``soprano's lyric trill'' \cite{scratch1989}. Importantly, for such a note to be produced,  the blade cannot be flat or bent into a `J-shape' (Fig.~1B), but must be bent into an `S-shape' (Fig.~1C). This geometric transformation allows the saw to sing  and is well-known to musicians who describe the presence of a `sweet spot', i.e., the inflection curve in the `S-shaped' blade; bowing near it produces the clearest notes, while bowing far from it causes the saw to fall silent \cite{scratch1989}. Early works \cite{cook,tubis1982}, including notably by Scott and Woodhouse \cite{ScottWoodhouse}, attempted to understand this peculiar feature by analyzing the linearized vibrational modes of a thin elastic shell \cite{koiter1960,audoly2010}. Through a simplified asymptotic analysis, they showed that a localized vibrational eigenmode emerges at an inflection point in a shell with spatially varying curvature, and is responsible for the musicality of the saw. Recent works have reproduced this result using numerical simulations \cite{worland2016musical,worland2019musical} but a deeper understanding of the origin of localization has remained elusive.

A simple demonstration of playing the saw quickly reveals the robustness of its musical quality to imperfections in the saw, irregularities in its shape and the precise details of how the blade is flexed. Fig.~1D shows a time trace and spectrogram of the saw clamped in either a J-shape or an S-shape (Fig.~1B,C) when struck or bowed near the sweet spot. The dull and short-lived sound (Audio 1) associated with the J-shape might be contrasted with the nearly pure tone ($\approx 595~$Hz) lasting several seconds (Audio 2) when the saw is bowed while shaped like an S. While the pitch can be varied by changing the curvature of the saw, the sustained quality of the note is largely indifferent to the manner of excitation and the specific nature of the clamps, as long as the inflection point is present. The lack of sensitivity to these details suggests a topological origin for the localized mode responsible for the saw's striking sonority. That topology can have implications for band structures and the presence of edge conducting states even when the bulk is insulating was originally explored in electronic aspects of condensed matter to explain the quantization of the Hall conductance \cite{thouless1982quantized}, and lead to the prediction of topological insulators \cite{hasan2010colloquium,qi2011topological}. More recently, similar ideas have been used to understand  the topological properties of mechanical excitations, e.g., acoustic and floppy modes in discrete periodic lattices \cite{kane2014topological,huber2016topological,mao2018maxwell,ma2019topological}, in continuum elasticity \cite{bartolo2019topological,saremi2020topological,sun2020continuum,sun2019maxwell}, in fluid dynamics in geophysical and active matter systems \cite{delplace2017topological,shankar2017topological,souslov2019topological,shankar2020topological} etc. In many of the aforementioned systems, it is the breaking of time-reversal symmetry that leads to the appearance of topologically protected modes. Here we expand the use of topological ideas to show that underlying the time-reversible Newtonian dynamics of the singing saw is a new topological invariant that characterizes the propagation of waves in thin shells, arising from the breaking of up-down inversion symmetry by curvature.

\begin{figure*}[]
	\centering{\includegraphics[width=0.95\textwidth]{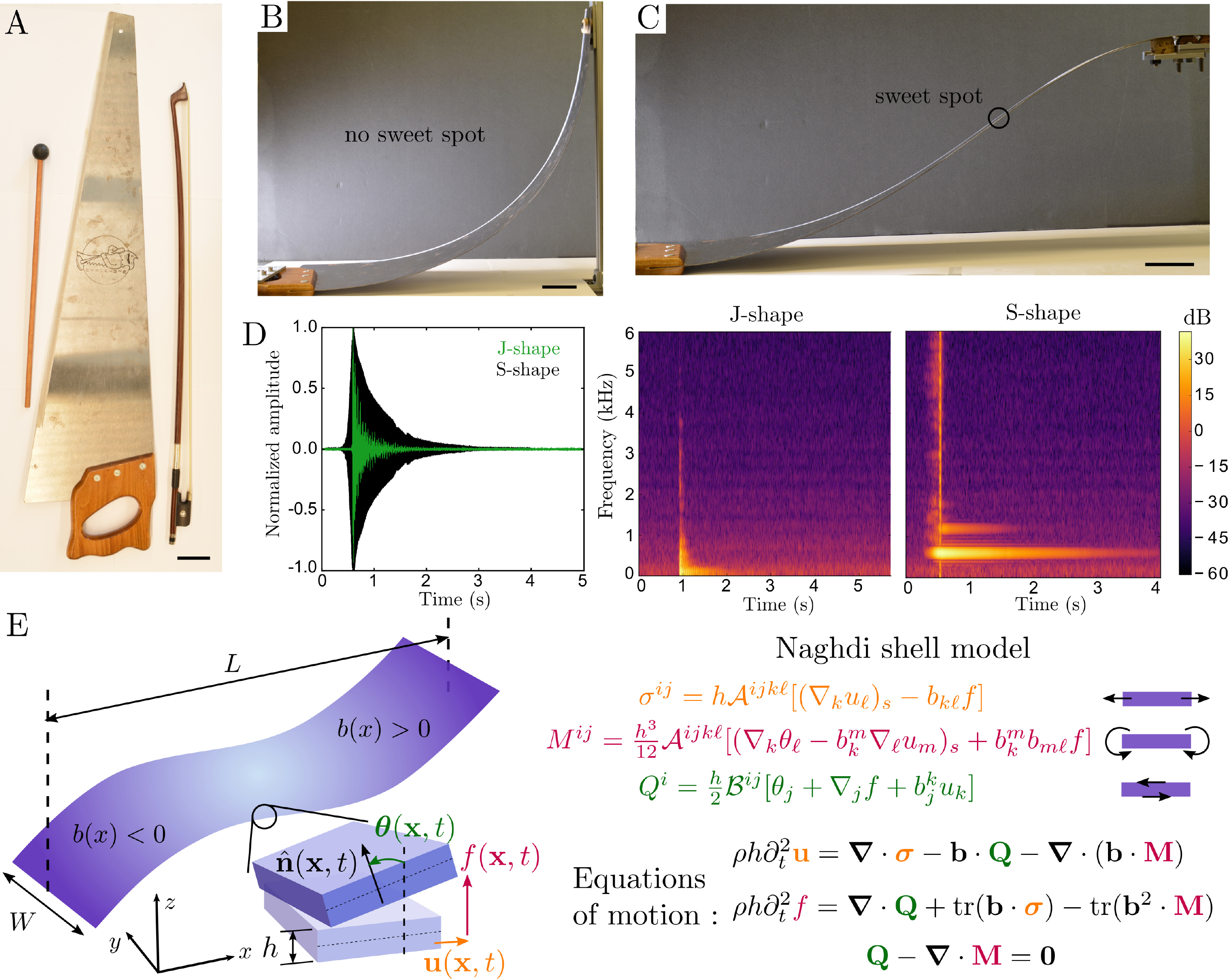}}
	\caption{{\bf The musical saw and its mathematical model.} (A) A violin bow and mallet placed alongside the saw. We clamp the saw in two configurations, (B) a J shape, and (C) an S-shape which is required to play music. The primary distinction between the two is that (C) has an inflection point (the `sweet spot') in its profile, while (B) has curvature of constant sign. Scale bar, $5~$cm. (D) Time series of the normalized audio signal when the saw in (B) is struck (green) and when the saw in (C) is bowed (black). The corresponding spectrograms for both the J-shape (B) and the S-shape (C) are shown on the right. The signal decays rapidly for the J-shape with a wider spread in frequency, while for the J-shape, a single dominant note with $\omega\approx595~$Hz survives the ringdown of the blade lasting several seconds. (E) A schematic of a blade of length $L$, width $W$ and thickness $h$ is sketched with a uniaxial curvature profile $b_{xx}(x)\equiv b(x)$ that changes sign along the $x$-axis as in (B). The saw can be modeled as an elastic shell whose deformations include an in-plane displacement $\b{u}$, a mid-surface deflection $f$ normal to the shell, and a rotation $\vec{\theta}$ of the local normal $\hat{\b{n}}$ as degrees of freedom ($\b{x}=(x,y)$ is the spatial coordinate). Elastic tensors $\mathcal{A}^{ijk\ell}$ and $\mathcal{B}^{ij}$ enter the constitutive equations (subscript `s' denotes symmetrization) for the in-plane stress ($\vec{\sigma}$), bending moment ($\b{M}$), and transverse shear ($\b{Q}$), see SI-Sec.~II. Derivatives are interpreted as covariant, and index manipulations employ the reference metric of the shell (SI-Sec.~II). The Kirchhoff limit for a shallow shell simplifies the dynamics to $\vec{\del}\cdot\vec{\sigma}=\b{0}$, $\rho h\partial_t^2f=\vec{\del}\vec{\del}:\b{M}+\tr(\b{b}\cdot\vec{\sigma})$ along with $\vec{\theta}=-\vec{\del}f-\b{b}\cdot\b{u}$ (see SI-Secs.~II,~III).}
\end{figure*}

The saw is modeled as a very thin rectangular elastic shell (thickness $h \ll W <L$, where $W,L$ are the width and length of the strip) made of a material with Young's modulus $Y$, Poisson ratio $\nu$ and density $\rho$ (Fig.~1E).  Its geometry is characterized by a spatially varying curvature tensor (second fundamental form) $\b{b}(\b{x})$, where $\b{x}=(x,y)$ is the spatial coordinate in the plane  As the saw is bent only along the (long) $x$-axis, $b_{xx}(x)\equiv b(x)$ is the sole non-vanishing curvature. To describe its dynamical response, we take advantage of its slenderness and treat the saw as a thin elastic shell that can be bent, stretched, sheared and twisted. Before moving to a computational model that accounts for these modes of deformation as well as real boundary conditions, to gain some insight into the problem and expose the topological nature of elastic waves, it is instructive to instead consider a simplified description valid for shallow shells with slowly varying curvature. 

In a thin shallow shell ($h|b_{ij}|\ll 1$), as bending is energetically cheaper than stretching \cite{rayleigh}, shear becomes negligible ($\b{Q}\approx\b{0}$, see Fig.~1E) and in-plane deformations propagate much more rapidly (at the speed of sound $c=\sqrt{Y/\rho}$) so that the depth-averaged stresses can be assumed to equilibrated, i.e., $\partial_j\sigma_{ij}=0$ \cite{ScottWoodhouse,evans2013reflection}. In this limit, using the solution of these equations in terms of the Airy stress function $\chi$ ($\sigma_{ij}=\mathcal{P}_{ij}\del^2\chi$, where $\mathcal{P}_{ij}=\delta_{ij}-\partial_i\partial_j/\del^2$ is a projection operator, see SI-Sec.~III), the in-plane geometric compatibility relation and the linearized dynamical equations for transverse motions can be written as \cite{koiter1960,ciarlet2000}
\begin{eqnarray}
	&\frac{1}{Yh}\del^4\chi=-\mathcal{P}_{ij}\del^2(b_{ij}f)\;,\label{eq:chi}\\
	&\rho h\partial_t^2f=-\kappa\del^4f+b_{ij}\mathcal{P}_{ij}\del^2\chi\;,\label{eq:f}
\end{eqnarray}
Here $f$ is the out-of-plane deflection of the shell (Fig.~1E) and the bending rigidity $\kappa=Yh^3/[12(1-\nu^2)]$. Crucially, in-plane and flexural (out-of-plane) modes remain geometrically coupled in the presence of curvature even in the linearized setting (Eqs.~\ref{eq:chi},~\ref{eq:f}). For a shell bent with constant curvature along the $x$-axis, i.e., a section of a uniform cylinder, $b(x)=b_0$ is a constant. In the bulk of the system, disregarding boundaries, we can Fourier transform Eqs.~\ref{eq:chi},~\ref{eq:f} using the solution ansatz $f=f_\b{q}e^{-i\omega t+i\b{q}\cdot\b{x}}$ to obtain the dispersion relation for flexural waves  to be $\omega_\pm(\b{q})=\pm\sqrt{(\kappa/\rho h)q^4+c^2b_0^2(q_y/q)^4}$ (Fig.~2A), where $q=|\b{q}|$. When $q_y=0$, i.e., the sheet is undeformed in the transverse direction, it remains developable (with generators that run parallel to the $y$ direction) and the bending waves are gapless, i.e., $\omega \rightarrow 0$ as $q \rightarrow 0$. However, when $q_y\neq 0$, a finite frequency gap $\sim c|b_0|$ controlled by the speed of sound and the curvature of the shell emerges as $q \rightarrow 0$ (Fig.~2A). Intuitively, this arises due to the geometric coupling between bending and stretching deformations in a curved shell which leads to an effective stiffening that forbids wave propagation below a frequency threshold. Similar spectral gaps appear in curved filaments and doubly curved shells as well \cite{evans2013reflection,kernes2021effects}. 

For the S-shaped saw, curvature scales of $b\sim 0.4-0.8~\rm{m}^{-1}$ are easily achievable (as in Fig.~1B-C), while the typical sound speed in steel is $c\sim5-6\times 10^3~$m/s so that the frequency gap is of order $2-5~$kHz. Comparing these estimates to the spectrogram in Fig.~1D (further quantified in Fig.~3) suggests that the localized mode excited upon bowing the S-shape saw (Fig.~1C) lies within the frequency gap. The J-shape saw (Fig.~1B) also vibrates at low frequencies (compared to the gap) when struck, presumably through the $q_y=0$ branch of delocalized flexural modes, though higher frequencies above the band gap can be excited by careful bowing (see SI-Fig.~S1A,B).

To unveil the topological structure of the vibration spectrum of the saw, we cast the second order dynamical equations (Eqs.~\ref{eq:chi},~\ref{eq:f}) in terms of first order equations by taking the ``square root'' of the dynamical matrix \cite{kane2014topological,susstrunk2016classification}. Focusing on the flexural modes alone, we obtain a Schr{\"o}dinger-like equation for the transverse deflections of a shallow shell (see SI-Sec.~III),
\begin{equation}
	\frac{i}{c}\partial_t\Psi=\mathcal{H}\Psi\;,\quad\mathcal{H}=\left(\matrix{
		0 & \mathcal{D}^\dagger \cr
		\mathcal{D} & 0} \right)
\end{equation}
where $\Psi=(c\mathcal{D}^\dagger f,\;i\partial_tf)$ and $\mathcal{D}=i\sqrt{\kappa/Yh}\del^2+b_{ij}\mathcal{P}_{ij}$ and $\dagger$ represents the conjugate transpose. The eigenvalues of the effective Hamiltonian $\mathcal{H}$ are given by the previously derived ($\omega_{\pm}(\b{q})$), and its complex eigenvectors $\Psi_\pm(\b{q})$ encode the topology of the band structure. The singularities in the arbitrary phase of the eigenvectors signals nontrivial band topology. To understand the phase of eigenvectors along the saw's long direction, we can consider fixing the transverse wavevector $q_y\neq 0$, leading to an effective one-dimensional (1D) system along the $x$-axis. Then the obstruction to continuously define the phase of the eigenvectors at every $q_x$ in Fourier space while respecting all the symmetries of the problem is quantified by the 1D Berry connection $\mathcal{A}(q_x)=i\sum_{n=\pm}\Psi_{n}(q_x)^\dagger\partial_{q_x}\Psi_{n}(q_x)$ (the $q_y$ dependence is suppressed) \cite{frankel2011geometry,nakahara2018geometry}. But what are the symmetries of our elastodynamic system?

One important symmetry is that imposed by classical time-reversal invariance in a passive, reciprocal material ($\mathcal{C}$: $x\to x$, $t\to-t$, $\Psi\to\Psi^*$, see SI-Sec.~III), which maps forward moving waves into backward moving ones, and guarantees that eigenmodes appear in complex-conjugate pairs \cite{susstrunk2016classification}.
%For the Hamiltonian in Eq.~3, time-reversal is implemented as $\mathcal{C}\mathcal{H}(q_x)\mathcal{C}^{-1}=-\mathcal{H}(-q_x)$ , with the operator $\mathcal{C}=i\vec{\tau}_yK$, where $\vec{\tau}_y$ is the second Pauli matrix and $K$ effects complex conjugation ($K^2=1$).
A second symmetry special to the saw is an emergent spatial reflection symmetry in the local tangent plane ($\Pi$: $x\to -x$, $t\to t$, $\Psi\to\Psi$, see SI-Sec.~III), which originates from the uniaxial nature of the prescribed curvature along the $x$ axis, and the insensitivity of bending to the orientation of the local tangent plane, a symmetry that is inherited from three dimensional (3D) rotational invariance.  Upon simultaneously enforcing both \emph{dynamical} and \emph{spatial} symmetries, a new topological obstruction posed by curvature emerges, and is quantified by a $\b{Z}_2$ index (see SI-Sec.~III)
\begin{equation}
	(-1)^{\nu}=\exp\left[i\int_0^\infty\dd q_x\;\mathcal{A}(q_x)\right]\frac{\rm{Pf}[\b{W}(0)]}{\rm{Pf}[\b{W}(\infty)]}\;,
\end{equation}
similar to topological insulators with crystalline symmetries \cite{fu2007topological,fu2011topological,hughes2011inversion}. $\rm{Pf}[\b{W}]$ denotes the Pfaffian of the antisymmetric overlap matrix $W_{ij}(q_x)=\Psi_i(q_x)^\dagger\mathcal{C}\Pi\Psi_j(q_x)$ ($i,j=\pm$). We note that unlike the mechanical Su-Schrieffer-Heeger chain \cite{kane2014topological} that exhibits a topological polarization in 1D, the emergent spatial reflection symmetry in our problem forces this polarization to vanish (see SI-Sec.~III).

As we work in the continuum, only differences in the topological invariant are well-defined independent of microscopic details. Across an interface at which curvature changes sign, i.e., a curvature domain wall, the topological invariant changes by $\Delta\nu=1$ (SI-Sec.~III). This directly demonstrates that the two oppositely curved sections of the saw behave as topologically nontrivial bulk systems that meet at the inflection line that functions as an internal ``edge''. As a result, nontrivial band topology underlies the emergence of the localized midgap mode, endowing it with robustness against details of the curvature profile and weakly nonlinear deformations (SI-Sec.~III).

\begin{figure*}[]
	\centering{\includegraphics[width=0.88\textwidth]{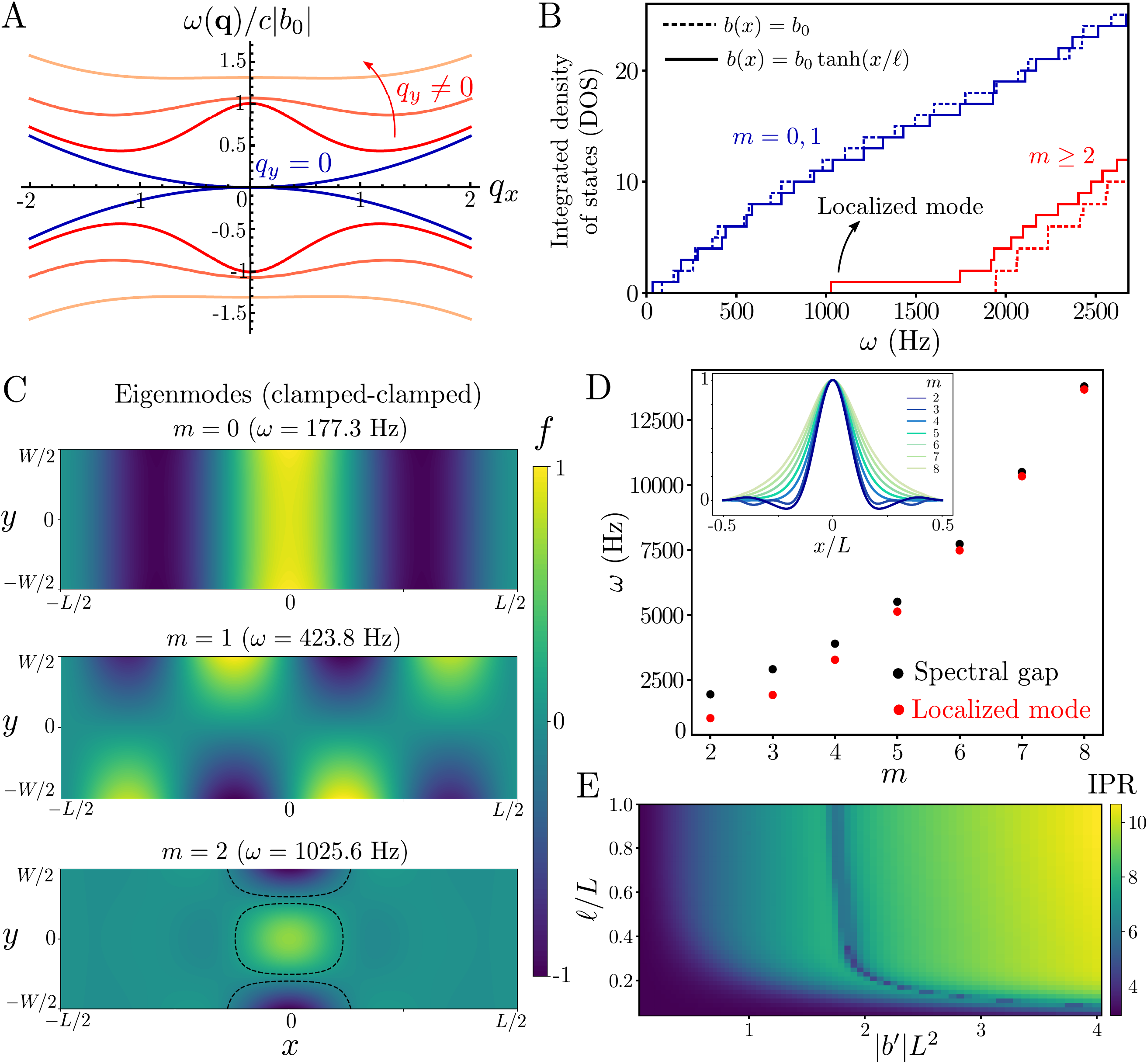}}
	\caption{{\bf Eigenmodes, band structure and topological localization.} (A) Analytical dispersion relation computed for an infinitely long strip with constant curvature along the $x$-axis ($h|b_0|=5\times10^{-5}$, $\nu=1/3$). The blue curves correspond to the $q_y=0$ gapless modes, and the red curves with $q_y\neq 0$ have a finite frequency gap. (B) Numerically computed integrated density of states for a finite curved strip ($b_0L=0.5$, $\ell/L=0.1$, $L=1~$m) with clamped-clamped boundary conditions. Developable eigenmodes (blue; labeled by discrete mode numbers $m=0,1$, akin to $q_y=0$ in the continuum) are gapless for both constant curvature (dashed) and the sigmoid profile (solid). Higher modes (red; $m\geq2$) exhibit a finite gap $\sim 2~$kHz for constant curvature (dashed), while the sigmoid profile features a localized mode ($\omega\sim1$~kHz) at the inflection point within the bulk band gap. (C) Numerical eigenmodes for the sigmoid profile with the local normalized deflection $f$ plotted (dashed lines are $10\%$ isocontours). Low frequency delocalized states with $m=0$ (top), $m=1$ (middle) and the first localized mode with $m=2$ (bottom). (D) Frequency of the localized modes (inset: normalized deflection at $y=W/2$ along $x$) and corresponding spectral gap for increasing transverse mode number $m\geq 2$. (E) Inverse participation ratio of the first localized mode for a piecewise linear curvature profile, plotted against the curvature gradient $b'$ and the length scale of curvature variation $\ell$.
	}
\end{figure*}
% Another reason: weak form admits solutions with less smoothness (see SI).
We test these predictions by numerically computing the eigenmodes of a finite elastic strip of length $L=1~$m, width $W=0.25~$m and thickness $h=10^{-3}~$m. For our shell model, we move away from the Kirchhoff model for shells and account for the kinematics associated with shear in addition to those associated with bending and stretching, as they effectively reduce the numerical ill-conditioning commonly seen in high-order continuum theories for slender plates and shells while allowing for numerical methods that require less smoothness and are easier to implement (see SI); together this allows for better computational accuracy. This framework forms the basis for the Naghdi shell-model \cite{naghdi1963} (see SI-Sec.~II for details) and accounts for an in-plane displacement vector along the mid-surface $\b{u}(\b{x},t)$, an out-of-plane deflection $f(\b{x},t)$ normal to the shell and an additional rotation $\vec{\theta}(\b{x},t)$ of the local normal itself (see Fig.~1E). These modes of deformations lead to depth-averaged stress resultants associated with stretching ($\vec{\sigma}$), bending ($\b{M}$) and shear ($\b{Q}$) as shown in Fig.~1E. The resulting covariant nonlinear shell theory along with inertial Newtonian dynamics provides an accurate and computationally tractable description of the elastodynamics of  thin shells (Fig.~1E and SI-Sec.~II). To highlight the topological robustness of our results, in our calculations we vary both the boundary conditions and curvature profiles imposed on the saw.

%We first choose asymmetric boundary conditions, with the left edge clamped, and the right edge free.
In Fig.~2B, the distribution of eigenmodes as a function of frequency is shown in the integrated density of states for a constant curvature shell, $b(x)=b_0$ (dashed lines), and an S-shaped shell with a smooth curvature profile $b(x)=b_0\tanh(x/\ell)$ (solid lines) that varies over a width $\ell$ near the inflection point at $x=0$. In both cases, the ends of the strip are kept clamped, and the spectra are calculated using an open-source code based on the finite element method \cite{fenics,fenicsshells}. As the curvature of the S-shape asymptotes to a constant $\pm b_0$ far from the origin, the bulk spectral gap and delocalized modes match that of the constant curvature case. Flexural modes that vary at most linearly in the $y$-direction (labeled by discrete mode numbers $m=0,1$, due to the lack of translational invariance) correspond to linearized isometries; they delocalize over the entire ribbon (Fig.~2C, top and middle) and populate states all the way to zero frequency, i.e., with a gapless spectrum. This is true for both constant curvature (dashed blue line, Fig.~2B) and the S-shape shell (solid blue line, Fig.~2B) as these bulk modes are unaffected by curvature. In contrast, all other modes that bend in both directions ($m\geq 2$) are generically gapped for a constant curvature profile (dashed red line, Fig.~2B) as expected. But for the S-shape, in addition to the gapped bulk modes, a new mode appears within the spectral gap (solid red line, Fig.~2B). This midgap state (shown here for $m=2$) is a localized mode that is trapped in the neighborhood of the inflection line (Fig.~2C, bottom). For increasing mode number $m\geq 2$, similar topological modes appear within the bulk bandgap, with growing localization lengths (Fig.~2D, inset) and higher frequencies (Fig.~2D), as predicted analytically (SI-Sec.~III). 
Qualitatively, the presence of an inflection line in the S-shaped saw makes it geometrically soft there;  the generators of cylindrical modes are now along the length of the saw, and the curved regions on either side that are geometrically stiff serve to insulate the soft internal "edge" from the real clamped edges.

Of particular note is that the localized modes, unlike the extended states, are virtually unaffected by the boundaries and the conditions there (see SI-Fig.~S2A for eigenmodes in a strip with asymmetric boundary conditions where the left edge is clamped and the right edge is free). Spatial gradients in curvature, however, \emph{do} impact the extent of localization. We demonstrate this using a piecewise continuous curvature profile that has a constant linear gradient $b'$ over a length $\ell$ across the origin and adopts a constant curvature outside this region. By varying both the curvature gradient $b'$ and the length scale $\ell$, we can tune the localization of the lowest topological mode (same as Fig.~2C, bottom), quantified by the inverse participation ratio ${\rm IPR}=\int\dd\b{x}|f(\b{x})|^4/(\int\dd\b{x}|f(\b{x})|^2)^2$ (Fig.~2E). Strong localization (high IPR) is quickly achieved for sharp gradients in curvature (${\rm IPR}\propto\sqrt{|b'|/h}$, see SI-Sec.~III) as long as the length scale of curvature variation is not too small ($\ell/L\geq0.1$, Fig.~2E), demonstrating the ease of geometric control of localization. 

\begin{figure*}[]
	\centering{\includegraphics[width=0.88\textwidth]{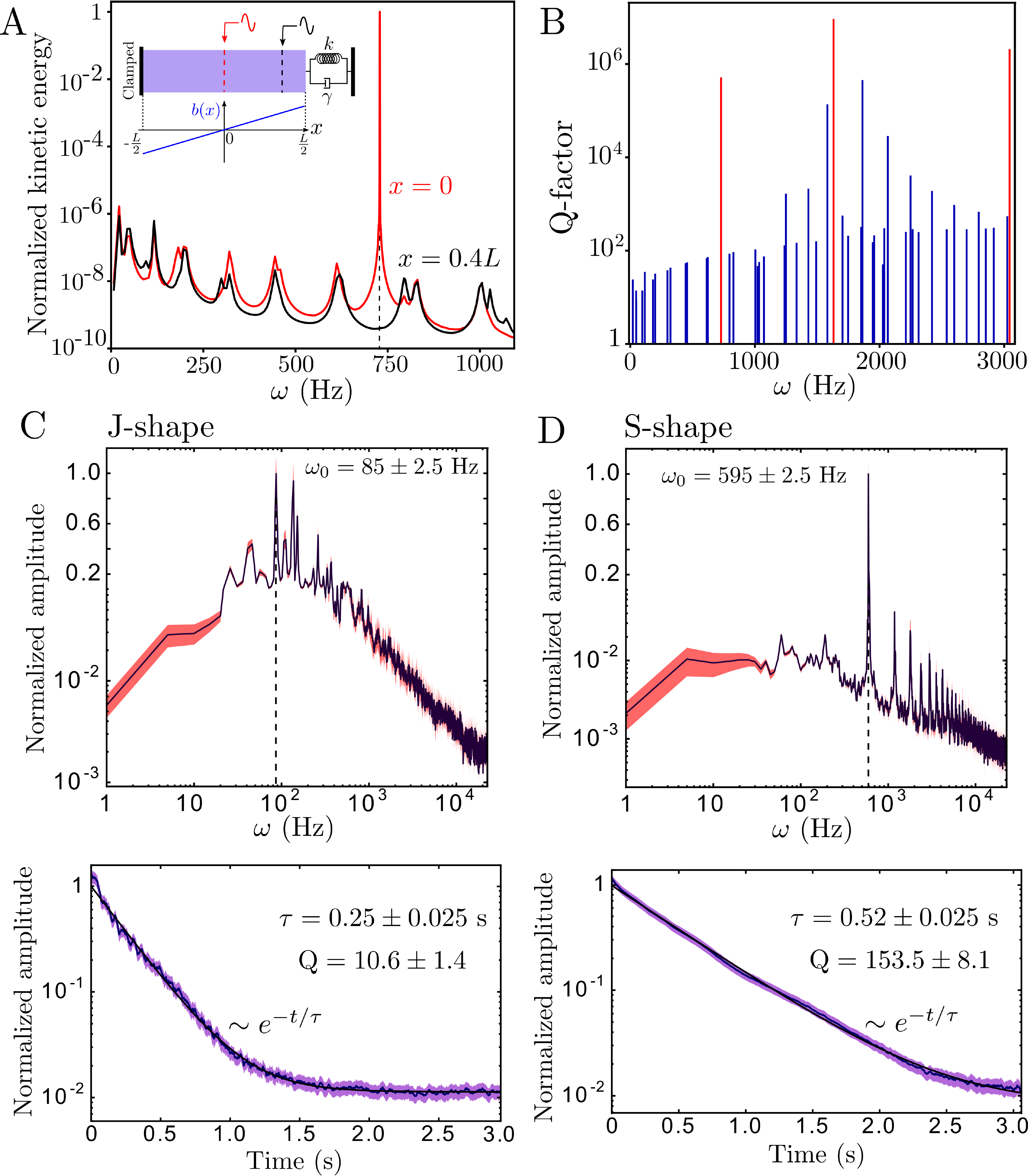}}
	\caption{{\bf Dissipative dynamics and high quality oscillators.} (A) Resonance curves for a shell with a linear curvature profile (inset) periodically driven at the inflection point ($x=0$, red) and away from it ($x=0.4L$, black) for varying frequency ($\omega\approx740~$Hz corresponds to the first localized mode). (B) Numerically computed Q-factor shows dramatic enhancement at localized mode frequencies (red) over delocalized modes (blue). (C-D) Experimental measurement of Q-factor (see SI-Sec.~I for details) for the musical saw in a (C) J-shape (Fig.~1B), and (D) S-shape (Fig.~1C). Note the normalized Fourier spectrum amplitude (C-D, top) is on a log scale below $0.1$ and linear above, with the peak frequency marked as $\omega_0$. The average signal decay (blue curve; C-D, bottom) is fit to a single decaying exponential (black curve; C-D, bottom). The shaded region is the standard error in both C-D.
	}
\end{figure*}

The boundary insensitivity of topologically localized modes has important dynamic consequences that can be harnessed to produce high quality resonators. The primary mode of dissipation in the saw, as in nanoelectromechanical devices \cite{wilson2011high}, is through substrate or anchoring losses at the boundary. Internal dissipation mechanisms (from e.g., plasticity, thermoelastic effects and radiation losses), though present, are considerably weaker and neglected here. To model dissipative dynamics, we retain clamped boundary conditions on the left end and augment the right boundary to include a restoring spring ($k$) and dissipative friction ($\gamma$) for both the in-plane forces and bending moments (see Fig.~3A, inset and SI-Sec.~II). Informed by Fig.~2E, we choose a linear curvature profile spanning the entire length of the shell to obtain a strongly-localized mode. Upon driving the shell into steady oscillations, with a periodic point force applied at the inflection point ($x=0$; Fig.~3A, red curve), we see an extremely sharp resonance peak right at the frequency of the first localized mode (Fig.~3A). In contrast, when the shell is driven closer to the boundary ($x=0.4L$; Fig.~3A, black curve), the response is atleast six orders of magnitude weaker as the localized mode isn't excited and only the delocalized modes contribute. Localization hence protects the mode from dissipative decay, unlike extended states that dampen rapidly through the boundaries. We further quantify this using a Q-factor computed from undriven relaxation of the shell initialized in a given eigenmode (SI-Sec.~II). Ultrahigh values of  ${\rm Q}\sim10^5-10^6$ are easily attained when a localized mode is excited (Fig.~3B, red), well over the Q-factor of all other modes (Fig.~3B, blue). Similar results are obtained for other curvature profiles as well, such as a sigmoid curve (SI-Fig.~S2B).

To compare these computational results with experiments , we perform ringdown measurements on a musical saw (see SI-Sec.~I for details) clamped in both the J-shape (Fig.~1B) and the S-shape (Fig.~1C). The normalized Fourier spectra and exponential decay ($\tau$) of the signal envelope are shown in Fig.~3C (J-shape) and Fig.~3D (S-shape) with the dominant frequency ($\omega_0$) marked. We find a factor $\sim15$ enhancement in the Q-factor (${\rm Q}=\omega_0\tau/2$) for the S-shape saw (${\rm Q}\sim150$; Fig.~3D, left) over the J-shape (${\rm Q}\sim10$; Fig.~3C, left). We emphasize that this significant quality factor improvement, though not as dramatic as the numerically computed Q-factors (Fig.~3B), is still striking given the initial impulse (mallet strike for J-shape, bow for S-shape, see SI-Fig.~S1 for other cases) excites an uncontrolled range of frequencies and other sources of energy loss including internal damping are presumably also present. 

Our combination of analysis, finite element simulations and experiments has demonstrated that a saw sings because its curvature generates a frequency gap in the acoustic spectrum which closes at an inflection point (line), that acts as an interior  ``edge'' allowing a localized mode to emerge within the band gap. Unlike mechanisms of weak localization \cite{heilman2010localized,filoche2012universal} or well-known ``whispering gallery modes'' \cite{rayleigh,rayleigh1910cxii} that rely sensitively on details of the domain geometry, our topological argument explains the existence of localized sound modes and their robustness against perturbations in the musical saw, providing a framework to explore not just topological mechanics, but also \emph{dynamics} in thin plates and shells. 

The ability to control spatial geometry to trap modes at ``interfaces'' in the interior of the system offers a unique opportunity to design high quality oscillators. As our results are material independent, they apply equally well to nanoscale electromechanical resonators \cite{craighead2000nanoelectromechanical,ekinci2005nanoelectromechanical}, and provide a geometric approach to design high quality resonators without relying on intrinsic nonlinearities \cite{lifshitz2008nonlinear}.
Just as in the musical saw, in nanomechanical devices dissipation can be dominated by radiation through the clamped boundary \cite{wilson2011high}.
%In most nanomechanical devices, as in the musical saw, dissipation dominates primarily at the anchored boundary through which vibrations leak energy and attenuate \cite{wilson2011high}.
Current on-chip topological nanoelectromechanical metamaterials use carefully patterned periodic arrays of nanomembranes to control localized modes in robust acoustic waveguides \cite{cha2018experimental,ma2021nanomechanical}. Our work suggests an alternate strategy inspired by the singing saw, that  relies solely on the scale separation intrinsic to any curved thin sheet; by manipulating curvature spatially,  topological modes localized in the interior hence remain vibrationally isolated and decay extremely slowly, allowing ultrahigh quality oscillations, perhaps even in the ultimate limit of atomically thin graphene \cite{bunch2007electromechanical}.  

%\bibliography{references}
%\bibliographystyle{Science}

\section*{Acknowledgments}
SS is supported by the Harvard Society of Fellows. SS would like to thank Vincenzo Vitelli for helpful discussions and gratefully acknowledges useful interactions during the virtual 2021 KITP program on ``The Physics of Elastic Films: from Biological Membranes to Extreme Mechanics'', supported in part by the National Science Foundation under Grant No. NSF PHY-1748958. LM acknowledges NSF DMR 2011754, NSF DMR 1922321, and the Henri Seydoux Fund for partial financial support.

%Here you should list the contents of your Supplementary Materials -- below is an example. 
%You should include a list of Supplementary figures, Tables, and any references that appear only in the SM. 
%Note that the reference numbering continues from the main text to the SM.
% In the example below, Refs. 4-10 were cited only in the SM.     
\section*{Supplementary materials}
Materials and Methods\\
Supplementary Text\\
Figs.~S1 to S3\\
%Tables S1 to S4\\
%References \textit{()}

\end{document}